\documentclass[12pt]{article}
\usepackage{amsmath,amssymb}
\usepackage{bm}
\usepackage{color}
\usepackage{cite}
\usepackage{mathtools}

\usepackage{multirow}

\begin{document}

\title{
\begin{flushright}
\ \\*[-80pt]
\begin{minipage}{0.2\linewidth}
\normalsize
EPHOU-22-009\\*[50pt]
\end{minipage}
\end{flushright}
{\Large \bf
Mass matrices with CP phase \\ in modular flavor symmetry
\\*[20pt]}}

\author{
Shota Kikuchi  $^{1}$,
~Tatsuo Kobayashi  $^{1}$, 
\\ Morimitsu Tanimoto $^{2}$,   and~Hikaru Uchida$^{1}$
\\*[20pt]
\centerline{
\begin{minipage}{\linewidth}
\begin{center}
$^1${\it \normalsize
Department of Physics, Hokkaido University, Sapporo 060-0810, Japan} \\*[5pt]
				$^2${\it \normalsize
Department of Physics, Niigata University, Niigata 950-2181, Japan} \\*[5pt]
\end{center}
\end{minipage}}
\\*[50pt]}

\date{
\centerline{\small \bf Abstract}
\begin{minipage}{0.9\linewidth}
\medskip
\medskip
\small
We study the CP violation and the CP phase of quark mass matrices in modular flavor symmetric models.
The CP symmetry remains at $\tau = e^{2 \pi i/3}$ by a combination of the $T$-symmetry of the modular symmetry.
However, the $T$-symmetry breaking may lead to the CP violation at  the fixed point $\tau = e^{2 \pi i/3}$.
We study such a possibility in magnetized orbifold models as examples of modular flavor symmetric models.
These models, in general,  have more than one candidates for Higgs modes, while 
generic string compactifications also lead to several Higgs modes.
These Higgs modes have different behaviors under the $T$-transformation.
The light Higgs mode can be a linear combination of those modes so as to lead to realistic quark mass matrices.
The CP phase  of mass matrix does not appear in a certain case, which is determined by 
the $T$-transformation behavior.  
Deviation from it is important to realize the physical CP phase.
We discuss an example leading to non-vanishing CP phase  at the fixed point $\tau = e^{2 \pi i/3}$.
\end{minipage}
}

\begin{titlepage}
\maketitle
\thispagestyle{empty}
\end{titlepage}

\newpage


\section{Introduction}
\label{Intro}

The origin of the flavor structure including the CP violation is one of important issues to study in particle physics.
The four-dimensional (4D) CP symmetry can be embedded into a proper Lorentz symmetry 
in higher dimensional theory such as superstring theory \cite{Green:1987mn,Strominger:1985it,Dine:1992ya,Choi:1992xp,Lim:1990bp,Kobayashi:1994ks}.
In such a theory, the CP symmetry can be broken spontaneously at the compactification scale 
or below it within the framework of 4D effective field theory.

In addition to the CP symmetry, 
geometrical symmetries of compact space 
can be sources of the flavor symmetries among quarks and leptons.
For example, $D_4$ and $\Delta(54)$ flavor symmetries can be 
derived from heterotic string theory on orbifolds and magnetized/intersecting 
D-brane models 
\cite{Kobayashi:2004ya,Kobayashi:2006wq,Ko:2007dz,Beye:2014nxa,Abe:2009vi}.
These non-Abelian discrete flavor symmetries have been used 
in model building for the quark and lepton flavors from the bottom-up approach 
\cite{
	Altarelli:2010gt,Ishimori:2010au,Ishimori:2012zz,Hernandez:2012ra,
	King:2013eh,King:2014nza,Tanimoto:2015nfa,King:2017guk,Petcov:2017ggy,Kobayashi:2022moq}
.

The torus $T^2$ and the orbifold $T^2/Z_2$ compactifications have the modular symmetry, which 
corresponds to the change of basis and is generated by $S$ and $T$ generators.
Interestingly, the modular symmetry transforms zero-modes of matter fields.
That is, the modular symmetry can also be a source of the flavor symmetry 
of quarks and leptons.
(See for heterotic string theory on orbifolds \cite{Ferrara:1989qb,Lerche:1989cs,Lauer:1990tm} and magnetized D-brane models \cite{Kobayashi:2018rad,Kobayashi:2018bff,Ohki:2020bpo,Kikuchi:2020frp,Kikuchi:2020nxn,
Kikuchi:2021ogn,Almumin:2021fbk}.\footnote{
Calabi-Yau compactifications include more moduli and they have larger 
symplectic modular symmetries \cite{Strominger:1990pd,Candelas:1990pi,Ishiguro:2020nuf,Ishiguro:2021ccl}.})

Inspired by these extra dimensional models and superstring theory, 
recently 4D modular flavor symmetric models have been studied in lepton and quark sectors.
Indeed, the well-known finite groups $S_3$, $A_4$, $S_4$ and $A_5$
are isomorphic to the finite modular groups 
$\Gamma_N$ for $N=2,3,4,5$, respectively\cite{deAdelhartToorop:2011re}.
The lepton mass matrices have been given successfully  in terms of {$\Gamma_3\simeq A_4$} modular forms \cite{Feruglio:2017spp}.
Modular invariant flavor models have also been proposed on the $\Gamma_2\simeq  S_3$ \cite{Kobayashi:2018vbk},
$\Gamma_4 \simeq  S_4$ \cite{Penedo:2018nmg} and  
$\Gamma_5 \simeq  A_5$ \cite{Novichkov:2018nkm,Ding:2019xna}.
Other finite groups are also derived from magnetized D-brane models \cite{Kobayashi:2018bff}.
By using these modular forms, 
phenomenological studies of the lepton flavors have been done intensively
based on  $A_4$ \cite{Criado:2018thu,Kobayashi:2018scp,Ding:2019zxk,Okada:2020brs,Yao:2020qyy} and 
$S_4$ \cite{Novichkov:2018ovf,Kobayashi:2019mna,Wang:2019ovr}.
The quark mass matrices also have been discussed in order to reproduce observed Cabibbo-Kobayahi-Maskawa (CKM) matrix elements.
\cite{Okada:2018yrn,Okada:2019uoy}.
 A lot of references of the mudular invariant flavor models are seen  in Ref.\cite{Kobayashi:2022moq}.


We denote the complex structure modulus $\tau$ on $T^2$ as well as its orbifolds.
The modulus $\tau$ transforms under the CP symmetry 
as 
\begin{align}
\tau \to -\tau^*,
\label{eq:CP-tau}
\end{align}
because the 4D CP symmetry can be embedded into 
a proper Lorentz transformation of higher dimensions.
Thus, the CP symmetry remains along ${\rm Re}~\tau =0$.
On the other hand, the modulus value ${\rm Re}~\tau=-1/2$ 
transforms to  ${\rm Re}~\tau=1/2$ under the above CP transformation.
However, these modulus values ${\rm Re}~\tau=\pm 1/2$
are equivalent in modular symmetric models by the $T$ transformation.
The CP symmetry remains along ${\rm Re}~\tau=\pm 1/2$.
This issue was studied explicitly in Ref.~\cite{Kobayashi:2019uyt}.
Hence, the CP symmetry and modular flavor symmetries are combined 
so as to construct a larger symmetry 
\cite{Baur:2019kwi,Novichkov:2019sqv,Baur:2019iai,Baur:2020jwc,Nilles:2020gvu,Ishiguro:2021ccl}.\footnote{See 
also for CP in Calabi-Yau compactification \cite{Bonisch:2022slo}.}
The CP can be violated at a generic value of the modulus $\tau$.
For example, realization of the Kobayashi-Maskawa CP phase as well as quark masses and mixing angles 
in magnetized orbifold models 
was studied in Ref.~\cite{Kobayashi:2016qag}.

Indeed, how to fix the modulus value is an important issue, that is, the moduli 
stabilization problem, 
although the modulus value is used as a free parameter in many modular flavor symmetric models.
The spontaneous CP violation has been studied 
through the moduli stabilization.
(See for early works Refs.~\cite{Acharya:1995ag,Dent:2001cc,Khalil:2001dr,Giedt:2002ns}.)
Recently the CP violation was studied through the moduli stabilization 
due to the three-form fluxes \cite{Kobayashi:2020uaj,Ishiguro:2020nuf}.
For example, the modulus stabilization analysis in Ref.~\cite{Ishiguro:2020tmo} shows 
that the modulus can be stabilized at the fixed point $\tau=\omega$ 
with a highest provability, where $\omega = e^{2\pi i/3}$.
(See 
also Ref.~\cite{Abe:2020vmv,Novichkov:2022wvg,Ishiguro:2022pde}.)
At the fixed point $\tau = \omega$ the residual symmetry $Z_3$ of the modular symmetry remains, 
while the residual $Z_2$ and $Z_4$  symmetries remain at the fixed point $\tau =i$ for $PSL(2,\mathbb{Z})$ and $SL(2,\mathbb{Z})$, 
respectively.
These fixed points are also attractive from the viewpoint of model building in the bottom-up approach.
The large flavor mixing angle is simply realized
 at $\tau =i$ \cite{Okada:2020ukr} due to the residual $Z_2$  symmetry.
  (See also \cite{King:2019vhv}.)
Interestingly, the hierarchy of charged lepton masses  are successfully obtained at nearby $\tau =\omega$ 
 without tuning parameters \cite{Novichkov:2021evw}.
(See also  \cite{Feruglio:2021dte}.)  The challenge to the quark sector is promising.
The residual symmetries at the fixed points are also useful to stabilize dark matter candidates  \cite{Kobayashi:2021ajl}.
Thus,  the modular flavor symmetric model  presents  the phenomenologically special feature at the fixed points.
More studies at the fixed points are required to solve the flavor problem such as the CP violation
as well as the mass hierarchy.


The CP is not violated at the fixed point $\tau = \omega$ through the above discussion.
That is, the CP symmetry is preserved at $\tau=\omega$ 
if the $T$ symmetry remains.
On the other hand, if $T$ symmetry is broken, the CP violation may occur 
at the fixed point $\tau=\omega$.
In generic string compactification, there are more than one candidates of Higgs modes, which have the same 
$SU(2)_L\times U(1)_Y$ quantum numbers and can couple with quarks and leptons.
The torus and orbifold compactifications with magnetic fluxes are 
interesting compactifications.
They can lead to 4D chiral theory, where the generation number is determined by 
magnetic fluxes \cite{Cremades:2004wa,Abe:2008fi,Abe:2013bca,Abe:2014noa}.
These magnetic fluxes determine the number of Higgs modes, which can couple with three generations of quarks and leptons 
\cite{Abe:2008sx,Abe:2015yva}.
Yukawa couplings are written by theta functions.
Realistic quark and leptons mass matrices were studied \cite{Abe:2012fj,Abe:2014vza,Fujimoto:2016zjs,Kikuchi:2021yog}.
In this paper, we show that such magnetized orbifold models with multi-Higgs modes can break the $T$-symmetry, 
and they can lead to the CP violation even at the fixed point $\tau=\omega$.

This paper is organized as follows.
In section 2, we give a brief review in the CP violation in modular flavor symmetric models, 
and then study the importance of the $T$-symmetry at the fixed point $\tau=\omega$.
In section 3, we study the CP violation in the quark sector of the magnetized orbifold models, which were studied in 
Ref.~\cite{Kikuchi:2021yog}.
In particular, the $T$-symmetry is broken and that leads to 
the CP violation  at the fixed point $\tau=\omega$.
Section 4 is our conclusion.


\section{CP}

Here, at first we give a review on the CP in modular flavor symmetric models, 
and then study one of key points in the CP violation within the framework of the modular flavor symmetry.
We focus on six-dimensional theory, that is, two extra dimensions.
Similarly, we can study ten-dimensional theory with six extra dimensions.

\subsection{CP symmetry}

Here, we briefly review on the CP in modular flavor symmetric models.
We use the complex coordinate $z=y_1+\tau y_2$ on two extra dimensions, where 
$y_1$ and $y_2$ are real coordinates and $\tau$ is the complex structure modulus.
On $T^2$, we identify $z \sim z +m+n\tau$, where $m$ and $n$ are integers.
We transform $z$ as $z \to z^*$ or $z \to -z^*$ at the same time as 
the 4D CP transformation.
Such a transformation corresponds to a six-dimensional proper Lorentz transformation.
Here, we use the latter transformation $z \to -z^*$, 
because it maps the upper half plane ${\rm Im}\tau > 0$ to the same half plane.
Then, the  modulus $\tau$ transforms as Eq.~(\ref{eq:CP-tau}) under this CP symmetry.
Obviously the line ${\rm Re \tau}=0$ is symmetric under Eq.~(\ref{eq:CP-tau}).
However, the CP symmetry seems to be violated at other points.
For example, the line ${\rm Re}\tau =-1/2$ transforms as 
\begin{align}
\tau =-1/2+i{\rm Im}\tau ~~\to~~ 1/2 + i{\rm Im} \tau,
\end{align}
and it is not invariant.

The modular symmetry transforms the modulus $\tau$ as 
\begin{align}
\tau ~\to~ \gamma \tau = \frac{a\tau + b}{c \tau + d},
\end{align}
by $SL(2,\mathbb{Z})$ element $\gamma$, 
\begin{align}
\gamma=
\begin{pmatrix}
a & b \\ c & d
\end{pmatrix},
\end{align}
where $a,b,c,d$ are integers satisfying $ad-bc=1$.
The modular symmetry is generated by two elements, $S$ and $T$,
\begin{align}
S:\tau ~\to~ -\frac{1}{\tau},\ \qquad~~ T:\tau ~\to~ \tau +1.
\end{align}

As mentioned above,  the line ${\rm Re}\tau =-1/2$ transforms to  the line ${\rm Re}\tau =1/2$ under Eq.~(\ref{eq:CP-tau}), 
and is not invariant.
However, these lines are transformed each other by the $T$-transformation.
Thus, the line ${\rm Re}\tau =-1/2$ is also CP-symmetric by combining the $T$-symmetry.

Here, we consider 4D supersymmetric effective theory derived from higher dimensional theory.\footnote{
See e.g. Ref.~\cite{Kikuchi:2022txy} and references therein.}
For example, we study the superpotential terms including quark Yukawa couplings,
\begin{align}
W(\tau)=Y^{(u)}_{ij\ell}(\tau)Q_iu_jH^u_\ell + Y^{(d)}_{ij\ell}(\tau)Q_id_jH^d_\ell,
\end{align}
 where $Q_i$, $u_j$, $d_j$ denote superfields corresponding to three generations of left-handed quarks, 
right-handed up-sector quarks, and right-handed down-sector quarks, respectively, 
and $H^{u,d}_\ell$ are up-sector and down-sector Higgs superfields.
Here we have added indexes for Higgs fields $H^{u,d}_\ell$, because 
string compactifications, in general, lead to more than one pairs of Higgs fields.
These quark and Higgs fields transform under the modular symmetry,
\begin{align}
\Phi_i ~\to~ \frac{1}{(c\tau + d)^{k_i}}\rho(\gamma)_{ij} \Phi_j,
\end{align}
where $-k_i$ denote the modular weights of 4D fields and $\rho(\gamma)_{ij}$ is a unitary matrix to 
represent the modular group.
The Yukawa couplings $ Y^{(u,d)}_{ij}(\tau)$ depend on the modulus $\tau$, and 
they are modular forms.
Similarly, we can study other terms in the superpotential.

The typical K\"ahler potential of the modulus field $\tau$ is written by 
\begin{align}
\hat K = -\ln (2 {\rm Im}\tau),
\end{align}
and tree-level K\"ahler potential of matter field with the modular weight $-k_i$ is 
written by 
\begin{align}
K_m= \frac{|\Phi_i|^2}{(2{\rm Im} \tau)^{k_i}}.
\label{eq:Kahler-matter}
\end{align}
The modular symmetry within the framework of supergravity theory 
requires $e^{\hat K}|W|^2$ to be invariant.

The supersymmetric models are CP-symmetric if $|W|^2$ is invariant, i.e.
\begin{align}
W(\tau) ~\to~ e^{i\chi}\overline{W(\tau)},
\end{align}
under the CP transformation with $\tau \to -\tau^*$ including the CP transformation of chiral matter fields.

\subsection{CP and $T$-symmetry}

Here, we study the implication of $T$-symmetry from the viewpoint of CP violation.
Suppose that the $T$-transformation is represented by $\rho(T)$ in our 4D effective field theory, and 
$\rho(T)$ satisfies
\begin{align}
\rho(T^N)=\mathbb{I},
\end{align}
that is, the $Z^{(T)}_N$ symmetry.
We consider the field basis such that the $T$ transformation is represented by diagonal 
matrices,
\begin{align}
T:Q_i ~\to~ e^{2\pi iP[{Q_i}]/N}Q_i,~~u_i ~\to~ e^{2\pi iP[{u_i}]/N}u_i,~~d_i ~\to~ e^{2\pi iP[{d_i]}/N}d_i,~~H^{u,d}_i ~\to~ e^{2\pi iP[{H^{u,d}_i}]/N}H^{u,d}_i,
\end{align}
where $P[\Phi_i]$ denote $Z^{(T)}_N$ charges of fields $\Phi_i$. 
The $T$-invariance of the superpotential requires the following $T$-transformation of 
Yukawa couplings:
\begin{align}
T:Y^{(u)}_{ij\ell}(\tau) ~\to~ e^{2\pi i P[{Y^u_{(i j \ell)}}]/N}Y^{(u)}_{ij\ell}(\tau) , \qquad 
Y^{(d)}_{ij\ell}(\tau) ~\to~ e^{2\pi iP[{Y^d_{(i j \ell)}}]/N}Y^{(d)}_{ij\ell}(\tau) ,
\end{align}
where 
\begin{align}
P[{Y^u_{(i j \ell)}}]= -(P[{Q_i}]+P[{u_j}] + P[{H^u_\ell}]), \qquad 
P[{Y^d_{(i j \ell)}]}= -(P[{Q_i}]+P[{d_j}] + P[{H^d_\ell}]).
\label{eq:Yukawa-Tcharge}
\end{align}
We use these $Z^{(T)}_N$ charges satisfying $0 \leq P[Y^d_{(i j \ell)}] < N$ 
and $0 \leq P[Y^u_{(i j \ell)}] < N$.

The Yukawa couplings, which are modular forms, can be expanded in terms of $q=e^{2\pi i \tau}$.
Since they satisfy the above $T$-transformation behavior, 
they can be written by 
\begin{align}
Y^{(u)}_{ij\ell}(\tau) =a_0q^{P[Y^u_{(i j \ell)}]/N} + a_1 qq^{P[Y^u_{(i j \ell)}]/N} +a_2 q^2q^{P[Y^u_{(i j \ell)}]/N}
+ \cdots 
=\tilde Y^{(u)}_{ij\ell}(q) q^{P[Y^u_{(i j \ell)}]/N}, \notag \\
Y^{(d)}_{ij\ell}(\tau) =b_0q^{P[Y^d_{(i j \ell)}]} + b_1 qq^{P[Y^d_{(i j \ell)}]/N} +b_2 q^2q^{P[Y^d_{(i j \ell)}]/N} 
+ \cdots 
=\tilde Y^{(d)}_{ij\ell}(q) q^{P[Y^u_{(i j \ell)}]/N},
\end{align}
where the functions $\tilde Y^{(u)}_{ij\ell}(q) $ and $\tilde Y^{(d)}_{ij\ell}(q) $ include 
only integer powers of $q$, i.e. $q^n$.

Here, let us consider the model with one pair of $H^u$ and $H^d$, which are 
$T$-invariant.
In addition, we set ${\rm Re} \tau = -1/2$.
In this model, the Yukawa couplings are written by 
\begin{align}
Y^{(u)}_{ij\ell}(\tau) 
=\hat Y^{(u)}_{ij}(q) e^{-\pi i P[Y^u_{(i j )}]/N}, \notag \\
Y^{(d)}_{ij\ell}(\tau) 
=\hat Y^{(d)}_{ij}(q) e^{-\pi i P[Y^d_{(i j )}]/N},
\end{align}
where we have omitted the indexes for Higgs fields, and 
\begin{align}
P[Y^u_{(i j )}]= -(P[{Q_i}]+P[{u_j}] ), \qquad 
P[Y^d_{(i j )}]= -(P[{Q_i}]+P[{d_j}] ),
\end{align}
\begin{align}
\hat Y^{(u)}_{ij}(q) = \tilde Y^{(u)}_{ij}(q) e^{-2 \pi P[Y^u_{(i j )}]{\rm Im}\tau/N} , \qquad 
\hat Y^{(d)}_{ij}(q) = \tilde Y^{(d)}_{ij}(q) e^{-2 \pi P[Y^d_{(i j )}]{\rm Im}\tau/N}.
\end{align}
The Yukawa couplings have phases $e^{-\pi iP[Y^{u,d}_{(i j )}]/N}$, although $\hat Y^{(u)}_{ij}(q) $ and $\hat Y^{(d)}_{ij}(q) $ are real.
However, these phases can be removed by the following rephasing of fields:
\begin{align}
Q_i' =e^{-\pi iP[{Q_i}]/N}Q_i, \quad u_i'= e^{-\pi iP[{u_i}]/N}u_i, \quad d'_i= e^{-\pi iP[{d_i}]/N}d_i.
\end{align}
Then, this model is CP-invariant.
In this discussion, the $T$-symmetry is important.

Similarly, we can study the model with one pair of $H^u$ and $H^d$, which transform 
non-trivially under the $T$-transformation
\begin{align}
T:H^{u,d} ~\to~ e^{2\pi iP[{H^{u,d}}]/N}H^{u,d}.
\end{align}
When these Higgs fields develop their vacuum expectation values (VEVs), 
mass matrices have phases, but those are overall phases, and not physical.
The Higgs VEVs break the $T$-symmetry and the $T$-symmetry is broken through the moduli stabilization at $\tau=\omega$, 
but those are not enough to realize the physical CP phase.

Unless the above structure is violated by any effects such as non-perturbative effects, 
the above discussion suggests that we need two or more Higgs VEV directions.
We denote them by 
\begin{align}
v^u_\ell =|v^u_\ell|e^{2\pi iP[v^u_\ell]/N}=\langle H^u_\ell \rangle, \qquad 
v^d_\ell =|v^d_\ell|e^{2\pi iP[v^d_\ell]/N}=\langle H^d_\ell \rangle,
\end{align}
where $P[v^u_\ell]$ or $P[v^d_\ell]$ is not integer for a generic VEV.
At any rate, the phase structure of mass matrices at $\tau = \omega$ is 
controlled by the $T$-symmetry.
If they satisfy 
\begin{align}
&-\frac12 \left(P[Y^{u}_{(ij\ell)}] + P[{Q_i}]+P[{u_j}] \right) + P[v^{u}_\ell]= {\rm constant~independent~of}~\ell, \notag \\
&-\frac12 \left(P[Y^{d}_{(ij\ell)}] + P[{Q_i}]+P[{d_j}] \right) + P[v^{d}_\ell]= {\rm constant~independent~of}~\ell,
\label{eq:CP-condition}
\end{align} 
in all of allowed Yukawa couplings  with $i,j$ fixed, one can cancel phase of mass matrix elements 
up to an overall phase 
by the $Z_N^{(T)}$ rotation.
We can compare this condition with the relations (\ref{eq:Yukawa-Tcharge}), where the factor $-1/2$ originates from 
${\rm Re}\tau=-1/2$.
We study this point in the next section by using magnetized orbifold models as an example.

This condition is also applied for the fixed point $\tau =i$.
The Yukawa couplings are real in our basis because of ${\rm Re}\tau=0$.
Then, the coefficients of $P[Y^{u}_{(ij\ell)}] + P[{Q_i}]+P[{u_j}]$ and $P[Y^{d}_{(ij\ell)}] + P[{Q_i}]+P[{d_j}]$ are zero.
That is, if we choose non-trivial phases of Higgs VEVs, which can not be removed by 
rephasing, the CP violation occurs.
For example, if all of the Higgs VEVs are real in our basis, 
the CP symmetry remain at the fixed point $\tau = i$.
However, the CP can be violated at the fixed point $\tau = \omega$, 
even if all of the Higgs VEVs are real in our basis.
Difference of Higgs VEV phases from their $Z_N^{(T)}$ charges are important.
The $Z_N^{(T)}$ charge pattern is the reference to judge whether 
the non-trivial CP phase appears or not.

\section{CP phase in magnetized orbifold models}

Here we study the CP phase derived from  magnetized orbifold models as an example of 
modular flavor symmetric models.

\subsection{Magnetized orbifold models}

First, we give a brief review on zero-mode wave functions on magnetized $T^2$ \cite{Cremades:2004wa}.
Higher dimensional fields, e.g. spinor field $\Psi(x,z)$, can be decomposed by
\begin{align}
\Psi(x,y)=\sum \chi_i(x) \psi_i(z)+ \cdots,
\end{align}
where $x$ denotes the 4D coordinate, the first term $\chi_i(x) \psi_i(z)$ corresponds to zero-modes, and 
the ellipsis denotes massive modes.
For simplicity, we explain them by use of  $U(1)$ theory.
We introduce $U(1)$ background magnetic flux,
\begin{align}
  F = dA = \frac{\pi iM}{{\rm Im}\tau} dz \wedge  d\bar{z}, \qquad   A = \frac{\pi M}{{\rm Im}\tau} {\rm Im} (\bar{z} dz),
\end{align}
where $M$ must be integer because of the Dirac quantization condition.
We consider the Dirac equation for the spinor with $U(1)$ charge $q=1$.
On $T^2$, the spinor $\psi$ has two components, $\psi=(\psi_{+},\psi_-)^T$.
For $M>0$, $\psi_+$ has $M$ zero-modes, but $\psi_-$ has no zero-modes.
On the other hand, for $M<0$, $\psi_-$ has $|M|$ zero-modes but $\psi_+$ has no zero-modes.
Thus, we can realize a chiral theory.
When $M>0$, the $j$-th zero-mode can be written by 
\begin{align}
  \psi^{j,|M|}_+ (z,\tau) = \left(\frac{|M|}{{\cal A}}\right)^{1/4} e^{i\pi |M|z \frac{{\rm Im}z}{{\rm Im}\tau}} \vartheta
  \begin{bmatrix}
    \frac{j}{|M|} \\ 0 \\
  \end{bmatrix}
  (|M|z,|M|\tau), \label{eq:wavefunction_T2} 
\end{align}
where ${\cal A}$ denotes the area of $T^2$ and $\vartheta$ denotes the Jacobi theta function defined by
\begin{align}
  \vartheta
  \begin{bmatrix}
    a \\ b \\
  \end{bmatrix}
  (\nu,\tau)
  =
  \sum_{\ell\in\mathbb{Z}} e^{\pi i(a+\ell)^2\tau} e^{2\pi i(a+\ell)(\nu+b)}.
\end{align}
Similarly, we can write zero-mode wave functions of $\psi_-$ for $M<0$.
Hereafter, we omit the chirality sign index $\pm$, but we denote the wave function by $\psi_{T^2}$.
Here we use the normalization, 
\begin{align}
  \int d^2z \psi^{i,|M|}_{T^2} (z,\tau) \left(\psi^{j,|M|}_{T^2}(z,\tau)\right)^* = (2{\rm Im}\tau)^{-1/2} \delta_{i,j}. \label{eq:normalization_torus} 
\end{align}
The Yukawa coupling of zero-modes is written by 
\begin{align}\label{eq:Yukawa}
  Y^{ijk} &= g \int d^2z ~\psi^{i,|M_1|}_{T^2} (z,\tau) ~\psi^{j,|M_2|}_{T^2} (z,\tau) ~\left(\psi^{k,|M_3|}_{T^2} (z,\tau)\right)^*  \\
  &
=g{\cal A}^{-1/2}\left|\frac{M_1M_2}{M_1+M_2}\right|^{1/4} \sum_m \delta_{k,i+j+|M_1M_2|m}
\vartheta
  \begin{bmatrix}
    \frac{|M_2|i-|M_1|j+|M_1M_2|m}{|M_1M_2(M_1+M_2)|} \\
    0 \\
  \end{bmatrix}
  (0,|M_1M_2(M_1+M_2)|, \notag
\end{align}
where $g$ is the 3-point coupling in higher dimensional theory.
The gauge invariance requires 
\begin{align}
\label{eq:selection-rule}
|M_1|+|M_2|=|M_3|,
\end{align}
for allowed Yukawa couplings.
Furthermore, the Kronecker delta $\delta_{k,i+j+|M_1M_2|m}$ implies the coupling selection rule among 
these modes.

The zero-mode wave functions transform under the $T$-symmetry as \cite{Kobayashi:2018rad,Kobayashi:2018bff,Ohki:2020bpo,Kikuchi:2020frp,Kikuchi:2020nxn,
Kikuchi:2021ogn,Almumin:2021fbk}
\begin{align}
T:\psi^{j,|M|}_{T^2}(z,\tau)~\to~e^{i \pi j^2/M} \psi^{j,|M|}_{T^2}(z,\tau),
\end{align}
when $M$ is even.\footnote{See for generic case Ref.~\cite{Kikuchi:2021ogn}.}
Thus, the $T$-transformation is represented by the diagonal matrix in 
this basis, and zero-modes have $Z^{(T)}_{2M}$ charges.
The zero-mode wave functions transform under the $S$-symmetry as
\begin{align}
S:\psi^{i,|M|}_{T^2}(z,\tau)~\to~(-\tau)^{1/2}\frac{e^{i \pi /4}}{\sqrt{|M|}}\sum_\ell e^{2\pi i  j \ell/M} \psi^{\ell,|M|}_{T^2}(z,\tau).
\end{align}
Note that the transformation of the 4D fields $\chi_i(x)$ is the inverse of $\psi_i(z)$ to make $\Psi(x,y)$ invariant.
For example, the 4D fields transform as 
\begin{align}
T:\chi^{j}(z)~\to~e^{-i \pi j^2/M} \chi^{j}(z),
\end{align}
under the $T$-transformation.

The $T^2/Z_2$ orbifold is constructed by identifying $z\sim -z$ on $T^2$.
Wave functions on $T^2/Z_2$ are classified into $Z_2$ even and odd modes,
\begin{align}
  \psi_{T^2/{Z}_2^m}(-z) = (-1)^m \psi_{T^2/{Z}_2^m} (z), 
\end{align}
where $m=0$ and 1 correspond to $Z_2$ even and odd modes.
The zero-mode wave functions in orbifold models with magnetic fluxes can be written by use of 
zero-mode wave functions on $T^2$ as \cite{Abe:2008fi,Abe:2013bca,Abe:2014noa}.
\begin{align}
  \psi_{T^2/{Z}_2^m}^{j,|M|}(z) &= {\cal N}^j \left(\psi_{T^2}^{j,|M|}(z)+(-1)^m\psi_{T^2}^{j,|M|}(-z)\right) \notag \\
  &= {\cal N}^j \left(\psi_{T^2}^{j,|M|}(z)+(-1)^m\psi_{T^2}^{|M|-j,|M|}(z)\right), \label{eq:zero-modesonorbifold}
\end{align}
where
\begin{align}
 {\cal N}^j = \left\{
  \begin{array}{l}
    1/2 \quad (j=0,|M|/2) \\
    1/\sqrt{2} \quad ({\rm otherwise})
  \end{array}
  \right..
\end{align}
Table \ref{tab:number} shows the number of zero-modes on magnetized $T^2/{Z}_2$ orbifold 
\cite{Abe:2008fi,Abe:2013bca,Abe:2014noa}. 
We can realize three generations by $Z_2$ even modes for $|M|=4,5$ and by 
$Z_2$ odd modes for $|M|=7,8$.
\begin{table}[h]
\centering
\begin{tabular}{|c|c|c|c|c|c|c|c|c|c|c|c|c|} \hline
$|M|$ & 1 & 2 & 3 & 4 & 5 & 6 & 7 & 8 & 9 & 10 & 11 & 12 \\ \hline
${Z}_2$-even & 1 & 2 & 2 & 3 & 3 & 4 & 4 & 5 & 5 & 6 & 6 & 7 \\ \hline
${Z}_2$-odd & 0 & 0 & 1 & 1 & 2 & 2 & 3 & 3 & 4 & 4 & 5 & 5 \\ \hline
\end{tabular}
\caption{The number of zero-modes on magnetized $T^2/{Z}_2$ orbifold.}
\label{tab:number}
\end{table}

Yukawa couplings in magnetized orbifold models can be calculated by
\begin{align}
  Y^{ijk} = g\int d^2z ~\psi_{T^2/{Z}_2^{m_1}}^{i,|M_1|} (z,\tau) ~\psi_{T^2/{Z}_2^{m_2}}^{j,|M_2|} (z,\tau) ~\left(\psi_{T^2/{Z}_2^{m_3}}^{k,|M_3|} (z,\tau)\right)^* .
\end{align}
Their explicit computations are straightforward by use of Eqs.~(\ref{eq:Yukawa}) and (\ref{eq:zero-modesonorbifold}).
Allowed Yukawa couplings must satisfy 
\begin{align}
\label{eq:selection-rule2}
m_1+m_2=m_3~~~ ({\rm mod}~~ 2),
\end{align}
 in addition to Eq.~(\ref{eq:selection-rule}).

\subsection{Quark mass matrix in magnetized orbifold models}

Here, we study quark mass matrices in the magnetized orbifold model, which was studied in 
Ref.~\cite{Kikuchi:2021yog}.
We consider the model that all of left-handed and right-handed quarks are 
originated from $Z_2$ even zero-modes with $M=4$, which lead to three zero-modes, that is, 
three generations.
(See for details of model building e.g. Refs.~\cite{Abe:2014vza,Abe:2017gye}.)
Because of Eqs.~(\ref{eq:selection-rule}) and (\ref{eq:selection-rule2}), Higgs modes correspond to 
$Z_2$ even zero-modes with $M=8$.
That means five pairs of $H^{u,d}_\ell$ from Table \ref{tab:number}.
These zero-mode wave functions are summarized in Table \ref{tab:three-model-448}.
\begin{table}[h]
\begin{center}
\renewcommand{\arraystretch}{1.2}
\begin{tabular}{c|c|c|c}
$i$ & $Q_i$ & $u_i, d_i$ & $H^{u,d}_i$ \\ \hline
0 & $\psi^{0,4}_{T^2}$ & $\psi^{0,4}_{T^2}$ & $\psi^{0,8}_{T^2}$ \\
1 & $\frac{1}{\sqrt{2}}(\psi^{1,4}_{T^2}+\psi^{3,4}_{T^2})$ & $\frac{1}{\sqrt{2}}(\psi^{1,4}_{T^2}+\psi^{3,4}_{T^2})$ & $\frac{1}{\sqrt{2}}(\psi^{1,8}_{T^2}+\psi^{7,8}_{T^2})$ \\
2 & $\psi^{2,4}_{T^2}$ & $\psi^{2,4}_{T^2}$ & $\frac{1}{\sqrt{2}}(\psi^{2,8}_{T^2}+\psi^{6,8}_{T^2})$ \\
3 & & & $\frac{1}{\sqrt{2}}(\psi^{3,8}_{T^2}+\psi^{5,8}_{T^2})$ \\
4 & & & $\psi^{4,8}_{T^2}$
\end{tabular}
\end{center}
\caption{Zero-mode wave functions.}
\label{tab:three-model-448}
\end{table}

Now, we study quark mass matrices in our model.
The mass matrices in the up and down sector of quarks can be written by
\begin{align}
(M_{u})_{ij}=\sum_\ell Y^{(u)}_{ij \ell} \langle H^u_\ell \rangle, \qquad (M_{d})_{ij}=\sum_\ell Y^{(d)}_{ij \ell} \langle H^d_\ell \rangle ,
\end{align}
when the Higgs fields develop their VEVs.
The Yukawa couplings can be written explicitly by
\begin{align}
  \begin{array}{ll}
    Y^{(u),(d)}_{ij0} = c\begin{pmatrix}
    X_0 &  &  \\
    & X_1 &  \\
    &  & X_2 \\
  \end{pmatrix}, &
  Y^{(u),(d)}_{ij1} = c\begin{pmatrix}
  & X_3 &  \\
  X_3 &  & X_4 \\
  & X_4 &  \\
  \end{pmatrix}, \\
  Y^{(u),(d)}_{ij2} = c\begin{pmatrix}
 &  & \sqrt{2}X_1 \\
 & \frac{1}{\sqrt{2}}(X_0 + X_2) &  \\
\sqrt{2}X_1 &  &  \\
\end{pmatrix}, &
  Y^{(u),(d)}_{ij3} = c\begin{pmatrix}
 & X_4 &  \\
X_4 &  & X_3 \\
 & X_3 &  \\
\end{pmatrix}, \\
  Y^{(u),(d)}_{ij4} = c\begin{pmatrix}
X_2 &  &  \\
 & X_1 &  \\
 &  & X_0 \\
\end{pmatrix},
& 
\end{array} \label{YukawaMatrix448}
\end{align}
where $c$ is an overall constant, and 
\begin{align}
  &X_0 = \eta_{0} + 2\eta_{32} + \eta_{64},  \qquad
  X_1 = \eta_{8} + \eta_{24} + \eta_{40} + \eta_{56},  \qquad
  X_2 = 2(\eta_{16} + \eta_{48}),  \notag \\
  &X_3 = \eta_{4} + \eta_{28} + \eta_{36} + \eta_{60}, \qquad 
  X_4 = \eta_{12} + \eta_{20} + \eta_{44} + \eta_{52}. 
\end{align}
Here, we have used the notation,
\begin{align}
  \eta_N = \vartheta
  \begin{bmatrix}
    \frac{N}{128} \\
    0 \\
  \end{bmatrix}
  (0,128\tau). 
\end{align}
Note that each of $Y^{(u),(d)}_{ij\ell}$ matrices with $\ell=0,1,2,3,4$ is not a rank-one matrix or 
an approximate rank-one matrix leading to the realistic quark mass hierarchy 
except ${\rm Im}\tau \to \infty$.
In addition, each   of $Y^{(u),(d)}_{ij\ell}$ matrices with $\ell=0,1,2,3,4$ has many zero elements,
which are originated from the coupling selection rule due to  $\delta_{k,i+j+|M_1M_2|m}$ in Eq.(\ref{eq:Yukawa}).
Therefore,  a single VEV direction is not realistic.

In particular, 
we set the modulus value as $\tau = \omega$ in order to study the quark mass matrices.
At this fixed point, the residual $Z_3$ symmetry, which is generated by $ST$, remains.
At the compactification energy scale, 
all of five pairs of Higgs fields are massless.
We expect that they generate their mass terms,
\begin{align}
\mu(\tau)_{ij} H^u_i H^d_j,
\end{align}
below the compactification scale.
Then, one pair remains light, and they develop their VEVs.
Such mass terms would be generated by non-perurbative effects such as D-brane instanton effects.\footnote{
Non-perturbative effects such as D-brane instanton effects may break some part of the modular symmetry \cite{Kikuchi:2022bkn}.}
Also coupling terms such as $Y(\tau)_{ij\ell} H^u_i H^d_j S_\ell$  may be origins of the 
mass terms when $S_\ell$ develop their VEVs like the next-to-minimal 
supersymmetric standard model.
Furthermore, higher order terms such as $Y(\tau)_{ij\ell_1\cdots \ell_n} H^u_i H^d_j S_{\ell_1}\cdots S_{\ell_n}$ 
may be their origins when $S_{\ell_i}$ develop their VEVs.
These depend on details of the model.
Thus, we take a phenomenological approach.
That is, we study which VEV directions lead to realistic results in quark mass matrices 
assuming such direction corresponds to the light mode in the above mass matrices.

Quark masses are hierarchical  very much.
That means that quark mass matrices are rank-one matrices  approximately, and 
realistic mass matrices deviate slightly from such rank-one matrices.
In Ref.~\cite{Kikuchi:2021yog}, VEV directions leading to rank-one mass matrices were 
studied.
We follow those  analysis.
For example, the $Z_3$ symmetry remains at the fixed point $\tau = \omega$.
In particular, VEV directions, which lead to rank-one mass matrices and $Z_3$ invariant vacuum, 
were studied in Ref.~\cite{Kikuchi:2021yog}.
Although the VEV directions in $Z_3$ eigenbasis shown in Ref.~\cite{Kikuchi:2021yog} include non-zero for only one $Z_3$ invariant direction and zeros for the other directions,~i.e. $A(1,0,0,0,0)$, the directions in our basis become
\begin{align}
&\langle H^u_\ell \rangle = \langle H^d_\ell \rangle = h_\ell, \\
&h_\ell =A(0.6254e^{0.04567i},0.6295e^{-0.1507i},0.2269e^{-0.7397i},0.04126e^{-1.721i},0.005421e^{-3.096i}).  \notag
\end{align}
It means that even if we consider this $Z_3$ invariant Higgs mode is the lightest  and only this Higgs field develop its real VEV, this direction is constructed by mixing of Higgs directions with different $Z^{(T)}_{N}$ charges, and each of VEV phases is different from 
its $Z^{(T)}_{N}$ charge.
Along this VEV direction, only the third generations gain masses, but the first and second generations are massless.
We take the following VEV directions
\begin{align}
 &\langle H_u^k \rangle = v_u(0.6228e^{0.1169i},0.6273e^{-0.1738i},0.2425e^{-1.055i},0.05774e^{-2.441i},0.01186e^{2.088i}), \notag \\
  &\langle H_d^k \rangle = v_d(0.6201e^{0.1713i},0.6259e^{-0.2009i},0.2605e^{-1.215i},0.06349e^{-2.538i},0.009710e^{1.930i}),
\label{eq:VEVs_ST}
\end{align}
which deviate slightly from the above rank-one directions.
When the above directions of Higgs modes are light and they develop 
their VEVs,  we realize the following quark mass matrices:
\begin{align}
  &M_u = m_t
  \begin{pmatrix}
    0.6228e^{0.1167i} & 0.4444e^{-0.3663i} & 0.08799e^{-1.840i} \\
    0.4444e^{-0.3663i} & 0.3219e^{-0.8600i} & 0.06626e^{-2.350i} \\
    0.08799e^{-1.840i} & 0.06626e^{-2.350i} & 0.01482e^{2.430i} \\
  \end{pmatrix}, \\
  &M_d = m_b
  \begin{pmatrix}
    0.6201e^{0.1712i} & 0.4433e^{-0.3926i} & 0.09452e^{-2.000i} \\
    0.4433e^{-0.3926i} & 0.3253e^{-0.9292i} & 0.06925e^{-2.438i} \\
    0.09452e^{-2.000i} & 0.06925e^{-2.438i} & 0.01194e^{2.388i} \\
  \end{pmatrix},
\end{align}
at $\tau = \omega$ in the orbifold wave function basis of Table \ref{tab:three-model-448}.
These mass matrices lead to the quark mass ratios,
\begin{align}
&\frac{m_u}{m_t}= 1.64 \times 10^{-5},\qquad \frac{m_c}{m_t}=6.22\times 10^{-3}, \notag \\
&\frac{m_d}{m_b}= 1.57 \times 10^{-3},\qquad \frac{m_s}{m_b}=1.32\times 10^{-2}.
\end{align}
These ratios are obtained at the compactification scale, which may be quite high.
When we compare them with experimental values, we have to evaluate renormalization group effects.
Renormalization group effects depend on 
the breaking scale of supersymmetry and $\tan \beta$.
For example, we compare them with mass ratios at the GUT scale by assuming the low-energy supersymmetric model
with $\tan \beta =5$ \cite{Bjorkeroth:2015ora}.
The Cabibbo-Kobayahi-Maskawa (CKM) matrix is also obtained in our model,
\begin{align}
|V_{\rm CKM}|=\begin{pmatrix}
        0.974 & 0.225 & 0.00405 \\
        0.225 & 0.974 & 0.0353 \\
        0.00719 & 0.0348 & 0.999 \\
      \end{pmatrix}.
\end{align}
Furthermore, our model leads to the Jarlskog invariant
\begin{align}
J=2.83\times 10^{-5}.
\end{align}
These results are shown in Table \ref{tab:MassandCKM_ST}.
Thus, our model can realize almost experimental values.
\begin{table}[h]
  \begin{center}
    \renewcommand{\arraystretch}{1.3}
    $\begin{array}{c|c|c} \hline
      & {\rm Obtained\ values} & {\rm Reference\ values} \\ \hline
      (m_u,m_c,m_t)/m_t & (1.64 \times 10^{-5},6.22\times 10^{-3},1) & (5.58\times 10^{-6},2.69\times 10^{-3},1) \\ \hline
      (m_d,m_s,m_b)/m_b & (1.57 \times 10^{-3},1.32\times 10^{-2},1) & (6.86\times 10^{-4},1.37\times 10^{-2},1) \\ \hline
      |V_{\rm CKM}| 
      &
      \begin{pmatrix}
        0.974 & 0.225 & 0.00405 \\
        0.225 & 0.974 & 0.0353 \\
        0.00719 & 0.0348 & 0.999 \\
      \end{pmatrix}
      & 
      \begin{pmatrix}
        0.974 & 0.227 & 0.00361 \\
        0.226 & 0.973 & 0.0405 \\
        0.00854 & 0.0398 & 0.999 
      \end{pmatrix}\\ \hline
      J_{CP} & 2.83 \times 10^{-5} & 2.80 \times 10^{-5} \\ \hline
    \end{array}$
  \caption{The mass ratios of the quarks, the  values of the CKM matrix elements, and Jarlskog invariant at $\tau=\omega$ under the vacuum alignments of Higgs fields in Eq.~(\ref{eq:VEVs_ST}).
    {Reference values} of mass ratios are shown in Ref \cite{Bjorkeroth:2015ora}.
    Ones of the CKM matrix elements and the Jarlskog invariant are shown in Ref \cite{Zyla:2020zbs}.}
    \label{tab:MassandCKM_ST}
  \end{center}
\end{table}


The important point in our results is that we can realize non-vanishing CP phase 
even at the fixed point $\tau = \omega$.
In section 2, it is found that the CP symmetry remains at the fixed  point $\tau = \omega$ 
because of the $T$-symmetry.
Now, let us investigate the $T$-symmetry in our model from the viewpoint of the CP violation.
At $\tau = \omega$, the Yukawa couplings have the following phases,
\begin{align}
X_0=|X_0|, \quad X_1=e^{-2\pi i/8}|X_1|, \qquad X_2 = -|X_2|, \notag \\
X_3=e^{-2\pi i/32}|X_3|, \qquad X_4=e^{-2\pi i 9/32}|X_4|. 
\end{align}
Also, the 4D quark fields $q_j=(Q_j,u_j,d_j)$ with $j=0,1,2$ in our model transform  
\begin{align}
T:q_j~\to~e^{-2\pi i j^2/8}q_j,
\end{align}
under $Z_N^{(T)}$, while
the Higgs modes transform as
\begin{align}
T: H^{u,d}_\ell~\to~e^{-2 \pi i \ell^2/16}H^{u,d}_\ell,
\end{align}
with $\ell =0,1,2,3,4$ under the $Z^{(T)}_{N}$ symmetry.
For simplicity, we consider the case that only the $H^{u,d}_0$ and $H^{u,d}_1$ develop their VEVs.
Then, the mass matrices can be written by 
\begin{align}
M_{u,d}=c
\begin{pmatrix}
   v_0|X_0| & v_1e^{-2 \pi i/32}|X_3| & 0 \\
 v_1e^{-2 \pi i/32}|X_3|    & v_0e^{-2\pi i/8}|X_1| & v_1e^{-2 \pi i9/32}|X_4|  \\
 0   &  v_1e^{-2 \pi i9/32}|X_4|  & -v_0|X_2| \\
  \end{pmatrix}.
\end{align}
When the phases of VEVs satisfy Eq.(\ref{eq:CP-condition}),~i.e.
\begin{align}
(v_0,v_1)=|a|e^{i\phi}(1,e^{-2\pi i/32}),
\end{align}
which are related to the $Z_N^{(T)}$ charges of $H^{u,d}_0$ and $H^{u,d}_1$, 
the phases in mass matrices can be canceled by the following rephasing of fields,
\begin{align}
q_j' =e^{-2\pi i j^2/16}q_j,
\end{align}
which are related to the $Z_N^{(T)}$ charges of $q_j$.
Similarly, we can discuss the case that more Higgs modes develop their VEVs.
What is important is the difference of the VEV phases from the $Z_N^{(T)}$ charge pattern.
On the other hand, 
if a single mode among $H^{u,d}_\ell$ develops its VEV, we can not realize the physical CP violation 
as discussed in section 2.
In the above example, it corresponds to $v_1=0$, and such a case leads to just an overall phase 
as discussed in section 2.
In order to realize non-vanishing CP phase, 
we need that a linear combination of more than one Higgs modes 
correspond to the light Higgs mode, and it develop its VEV, where the VEV phases must be different from 
the $Z_N^{(T)}$ charge pattern.
If VEV phases coincide with the $Z_N^{(T)}$ phase pattern, 
the physical CP does not appear.
Thus, the $Z_N^{(T)}$ phase pattern is the reference for the physical CP phase.
Note that we need such a linear combination to realize the realistic mass matrices in our model, 
which deviate slightly from the rank-one mass matrices.
The VEV of single mode can not lead to the realistic mass matrices.
Phenomenologically, we need Higgs modes 
along a generic VEV direction leading to rank-one mass matrices and 
the slight deviation from rank-one mass matrices.
When we require such a direction by phenomenological purpose, 
we can automatically realize the CP violation at the fixed point $\tau = \omega$ for 
VEV phases different from the $Z_N^{(T)}$ charge pattern.
This is an interesting scenario of the CP violation in modular flavor symmetric models.

Also, we comment an obvious example.
If the quark mass matrices are diagonal, we can always remove phases by $U(1)^9$ rotation, 
which may be independent of the $Z_N^{(T)}$ rotation.
Such an accidental symmetry may forbid the physical CP phase.

Also we show the example with the real VEV direction in our field basis.
We take the following VEV directions
\begin{align}
  &\langle H^u_\ell \rangle = v_u(0.7911, -0.6040, 0.09692, -0.00009578, -0.0002417), \notag \\
  &\langle H^d_\ell \rangle = v_d(0.7471,-0.6512,0.1333,-0.001655,-0.004176).
\label{eq:VEVs_real}
\end{align}
When the above directions of Higgs modes are light and they develop 
their VEVs,  we realize the following quark mass matrices:
\begin{align}
  &M_u = m_t
 \begin{pmatrix}
    0.7675&0.4170e^{0.9375\pi i}&0.034118e^{-\pi i/4}\\
    0.4170e^{0.9375\pi i}&0.2479e^{-0.1898\pi i}&0.02733e^{0.4383\pi i}\\
    0.034118e^{-\pi i/4}&0.02733e^{0.4383\pi i}&-0.006886\\
  \end{pmatrix}, \\
  &M_d = m_b
  \begin{pmatrix}
    0.7247&0.4495e^{0.9374\pi i}&0.04691e^{-\pi i/4}\\
    0.4495e^{0.9374\pi i}&0.2571e^{-0.1698\pi i}&0.02949e^{0.4498\pi i}\\
    0.04691e^{-\pi i/4}&0.02949e^{0.4498\pi i}&-0.01033\\
  \end{pmatrix},
\end{align}
at $\tau = \omega$ in the orbifold wave function basis of Table \ref{tab:three-model-448}.
These mass matrices lead to mass rations, mixing angles and the Jarlskog invariant 
shown in Table \ref{tab:MassandCKM_real}.
Also, this VEV direction  can realize almost experimental values except the ratio $m_u/m_t$.
\begin{table}[h]
  \begin{center}
    \renewcommand{\arraystretch}{1.3}
    $\begin{array}{c|c|c} \hline
      & {\rm Obtained\ values} & {\rm Reference\ values} \\ \hline
      (m_u,m_c,m_t)/m_t & (5.34 \times 10^{-5},4.68\times 10^{-2},1) & (5.58\times 10^{-6},2.69\times 10^{-3},1) \\ \hline
      (m_d,m_s,m_b)/m_b & (1.38 \times 10^{-3},4.10\times 10^{-2},1) & (6.86\times 10^{-4},1.37\times 10^{-2},1) \\ \hline
      |V_{\rm CKM}| 
      &
      \begin{pmatrix}
        0.975 & 0.223 & 0.00323 \\
        0.223 & 0.974 & 0.0410 \\
        0.0103 & 0.0398 & 0.999 \\
      \end{pmatrix}
      & 
      \begin{pmatrix}
        0.974 & 0.227 & 0.00361 \\
        0.226 & 0.973 & 0.0405 \\
        0.00854 & 0.0398 & 0.999 
      \end{pmatrix}\\ \hline
      J & 2.80 \times 10^{-5} & 2.80 \times 10^{-5} \\ \hline
    \end{array}$
  \caption{The mass ratios of the quarks, the  values of the CKM matrix elements, and Jarlskog invariant at $\tau=\omega$ under the vacuum alignments of Higgs fields in Eq.~(\ref{eq:VEVs_ST}).
    Reference values of mass ratios are shown in Ref \cite{Bjorkeroth:2015ora}.
    Ones of the CKM matrix elements and the Jarlskog invariant are shown in Ref \cite{Zyla:2020zbs}.}
    \label{tab:MassandCKM_real}
  \end{center}
\end{table}


We have concentrated on the modulus value ${\rm Re} \tau = \pm 1/2$, in particular 
the fixed point $\tau = \omega$.
This fixed point has the highest provability in the moduli stabilization analysis \cite{Ishiguro:2020tmo}, 
and is phenomenological interesting.
The next favorable values in the moduli stabilization analysis \cite{Ishiguro:2020tmo} are 
${\rm Re}\tau = \pm 1/4$ and $0$.
Even at ${\rm Re}\tau =  -1/4$, the phase structure is controlled by the $T$-symmetry.
The Yukawa couplings at this point have the following phase,
\begin{align}
X_0=|X_0|, \quad X_1=e^{-2\pi i/16}|X_1|, \qquad X_2 = -i|X_2|, \notag \\
X_3=e^{-2\pi i/64}|X_3|, \qquad X_4=e^{-2\pi i 9/64}|X_4|. 
\end{align}
Again for simplicity, we consider the case that only the $H^{u,d}_0$ and $H^{u,d}_1$ develop their VEVs.
\begin{align}
M_{u,d}=c
\begin{pmatrix}
   v_0|X_0| & v_1e^{-2 \pi i/64}|X_3| & 0 \\
 v_1e^{-2 \pi i/64}|X_3|    & v_0e^{-2\pi i/16}|X_1| & v_1e^{-2 \pi i9/64}|X_4|  \\
 0   &  v_1e^{-2 \pi i9/64}|X_4|  & -iv_0|X_2| \\
  \end{pmatrix}.
\end{align}
When the phases of VEVs satisfy 
\begin{align}
(v_0,v_1)=|a|e^{i\phi}(1,e^{-2\pi i/64}),
\end{align}
which are related to the $Z_N^{(T)}$ charges of $H^{u,d}_0$ and $H^{u,d}_1$, 
the phases in mass matrices can be canceled by the $Z_N^{(T)}$ rotation.
Thus, the $Z_N^{(T)}$ charge pattern is the reference to judge whether 
the non-trivial CP phase appears or not for ${\rm Re}\tau = \pm 1/4$, too.

One can expend our discussions for the modulus value ${\rm Re} \tau = \pm 1/n$, 
although we may not have a clear motivation to set ${\rm Re} \tau = \pm 1/n$ from the 
viewpoint of the moduli stabilization or phenomenology.
Also, the CP violation occurs at the $\tau=i$ along the Higgs VEV directions, where 
VEVs have relatively different phases.

\section{Conclusion}
\label{sec:conclusion}

We have studied the CP phase of quark mass matrices in modular flavor symmetric models at the fixed point of $\tau$.
The CP symmetry remains at ${\rm Re \tau}=\pm 1/2$, although 
${\rm Re \tau}=-1/2$ transforms to ${\rm Re \tau}=1/2$ under CP.
Its reason is that these transform each other under the $T$-symmetry.
That may suggest that if the $T$-symmetry is broken, 
the CP is also violated.
However, a simple breaking of $T$-symmetry is not enough to lead to the 
CP violation.

We have studied quark mass matrices in magnetized orbifold models.
Our model has five pairs of Higgs fields.
In general, string compactification leads to more than one candidates of Higgs modes.
We have computed quark mass matrices at the fixed point $\tau = \omega$.
This point is favorable from the viewpoint of moduli stabilization and also 
phenomenologically interesting in the flavor physics.
The CP is not violated at the fixed point $\tau = \omega$ 
if the $T$-symmetry remains.
In our model, non-vanishing physical CP phase can be realized.
The important point is that Higgs modes   mix each other.
Such mixing is required by the phenomenological purpose to realize realistic quark mass matrices, which are 
the approximate rank-one mass matrices.
The physical CP phase can appear when the phases of VEVs differ from the $Z_N^{(T)}$ charge.

We have shown a scenario to realize the CP violation in modular flavor symmetric models.
It is interesting to apply this scenario to other modular flavor symmetric models including the lepton sector, 
e.g. at the fixed point $\tau = \omega$.
By the phenomenological purpose, we required that the light Higgs modes correspond to 
linear combinations of Higgs modes.
It is important to show this point by computation of the $\mu$ mass matrices theoretically.
However, that is beyond our scope.
We would study it elsewhere.

We have found that the breaking of $T$-symmetry is important, and shown a scenario of $T$-symmetry breaking 
leading to the CP violation.
It would also be important to study whether another way to break the $T$-symmetry leading to 
the CP violation.



\vspace{1.5 cm}
\noindent
{\large\bf Acknowledgement}\\

This work was supported by  JSPS KAKENHI Grant Numbers JP22J10172 (SK), and JP20J20388 (HU).




\end{document}